\documentclass[superscriptaddress,amsmath,amssymb,aps,prl,twocolumn]{revtex4-1}
\pdfoutput=1
\usepackage{graphicx}
\usepackage{upgreek}
\newcommand{\ket}[1]{{\left| {#1} \right\rangle}}
\newcommand{\bra}[1]{{\left\langle {#1} \right|}}

\newcommand{\rme}{\mathrm{e}}

\newcommand{\ii}{\textsl{i}\,}
\newcommand{\sii}{\mbox{\scriptsize$\ii\!\!\:$\normalsize}}
\newcommand{\boldgreek}[1]{{\mbox{\boldmath$ {#1} $}}}

\begin{document}

\title{Certified quantum gates}

\author{Wesley C. Campbell}
\affiliation{University of California Los Angeles}

\date{\today}

\begin{abstract}
High quality, fully-programmable quantum processors are available with
small numbers ($<1000$) of qubits, and the scientific potential of
these near term machines is not well understood. If the small number
of physical qubits precludes practical quantum error correction, how
can these error-susceptible processors be used to perform useful
tasks? We present a strategy for developing quantum error detection
for certain gate imperfections that utilizes additional internal
states and does not require additional physical qubits. Examples for
adding error detection are provided for a universal gate set in the
trapped ion platform.  Error detection can be used to certify
individual gate operations against certain errors, and the
irreversible nature of the detection allows a result of a complex
computation to be checked at the end for error flags.

\end{abstract}

\maketitle

For the near-term future, it is likely that the quantum information
processors that become available will be capable of running
intermediate scale algorithms in the presence of multiple (possibly
numerous) errors \cite{Preskill2018quantum}.  For beyond-classical
computations this paradigm, the result reported by the quantum
computer is almost guaranteed to be wrong, and the recent observation
of quantum advantage by the Google group \cite{Google2019Supremacy}
was made possible only by arguing that after repeating the algorithm
many times, the algorithmic error probability could be made
statistically distinguishable from 1. For algorithms where the result
can be tested directly for correctness (such as Shor's factoring
algorithm \cite{Shor1997PolynomialTime}), this may be useful, at least
up to the point where the ratio of the run time to success probability
exceeds practical timescales. However, for many applications of
quantum computers (such as the sampling problem used to demonstrate
quantum advantage, and much of quantum simulation), the user has very
little idea which results are the trustworthy ones, potentially
rendering any purported quantum advantage effectively useless. Quantum
advantage is likely necessary, but not generally sufficient, to
realize quantum utility beyond classical machines.

Here, we consider the issue of how to deal with errors in quantum
processors caused by imperfections in the applied gates. While the
techniques we outline below are applicable to other hardware
platforms, we present them in the context of trapped ion hyperfine
qubits, which are effectively free of errors outside of those caused
by the gates themselves. In particular, since frequency stability is
typically easier to distribute, assess, and achieve than amplitude
stability, errors caused by frequency drifts are usually unlikely
compared to errors in the areas of pulses applied to perform gates,
and we therefore focus primarily on amplitude errors. Composite pulse
sequences \cite{Chuang2005NMR} can be used to suppress amplitude (and
frequency) errors that are common mode for the duration of a composite
pulse sequence, but do not perform well against correlated errors that
are not constant during the sequence, such as amplitude drifts from
amplifier temperature changes or laser intensity noise.  Far from
being exotic or implausibly insidious, these types of amplitude drift
errors, which degrade the protection afforded by composite pulse
sequences, have posed obstacles for a number of experiments working at
the forefront of fidelity
\cite{Blume-Kohout2017Demonstration,Brown2011SingleQubitGate,christensen2019high}.

In this paper, we present a strategy for designing certifiable gates
that uses  uxiliary states in each qubit host and does not require
additional physical qubits.  The larger Hilbert space afforded by
including ancillary states allows us to re-structure a gate as a
series of population transfer steps that are each followed by
dissipation of the error state through coupling to a bath/detector.
Specifically, each step is designed as a rotation from the initial
state $\ket{\psi_n}$ to an orthogonal target state $\ket{\psi_{n+1}}$.
By choosing $\ket{\psi_n}$ and $\ket{\psi_{n+1}}$ to reside in
orthogonal Hilbert spaces, this rotation can be attempted and
certified without acquiring knowledge of the information encoded in
either state. If the execution of this rotation is imperfect due to an
error in the degree of the rotation (i.e.~the amplitude), the system
will be left in $\ket{\phi_{n+1}} \approx \ket{\psi_{n+1}} + \epsilon
\ket{\psi_n}$, and subsequent detection that the system is not in the
error state $\ket{\psi_n}$ certifies the step against the rotation
error.  Since the dissipative detection step is irreversible, testing
for errors can be done either during the computation or at the end,
and checking a result for error flags can serve as a limited test of
the trustworthiness of the result.

We begin with an example that illustrates the main idea in the form of
arbitrary single-qubit gates that the user can certify against
single-pulse amplitude errors. Examples of how to certify against
errors in multi qubit contexts such as addressing errors and 2-qubit
entangling gate errors are also presented, demonstrating that a
complete set of gates for universal quantum computing can be augmented
with certification against some classes of errors.

\begin{figure}
\includegraphics[width=0.45\textwidth]{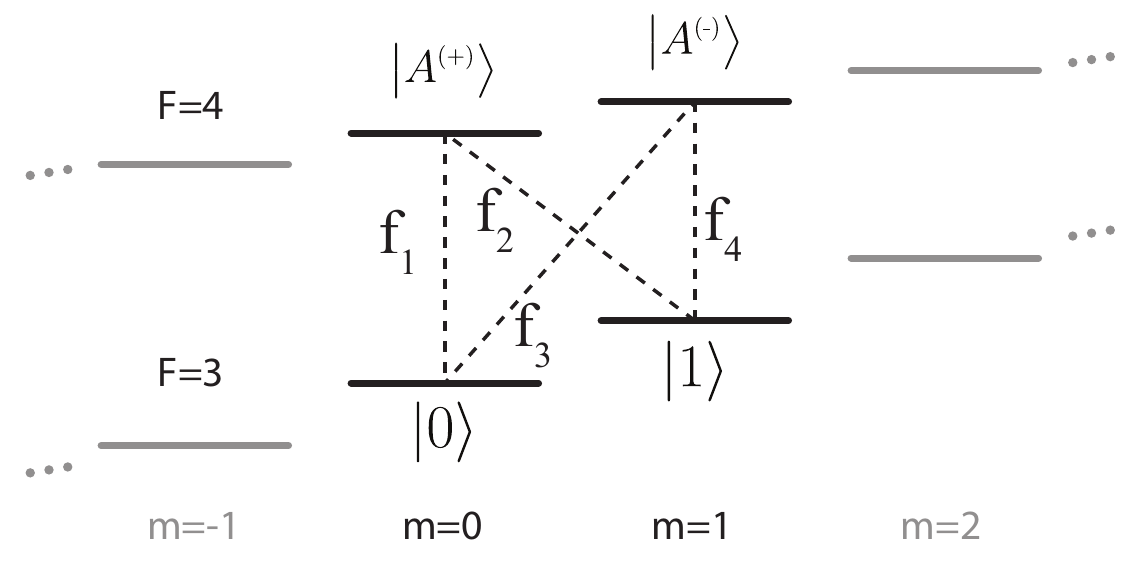}
\caption{Example ${}^2\!\!\,F_{7/2}^o$-state encoding of the qubit and
  auxiliary states in ${}^{171}\mathrm{Yb}^+$.  For storage, the qubit
  can be encoded in the two clock states and then transferred to and
  from this arrangement before and after gates.  All four transitions
  are within Zeeman shifts of the zero-field hyperfine splitting of
  $3.602(2)\mbox{ GHz}$ \cite{Taylor1999Measurement}.}
\label{fig:FStateZoom}
\end{figure}

We consider a system consisting of a qubit ($\ket{0}$ and $\ket{1}$)
and two additional long-lived auxiliary states $|A^{\mbox{\tiny(+)}}
\rangle$ and $|A^{\mbox{\tiny(-)}} \rangle$ that can each be coupled
to both qubit states via resonant radiation.  For concreteness, we
will suppose that the qubit and auxiliary states are encoded in Zeeman
sub-levels of the effectively stable ${}^2\!\!\,F_{7/2}^o$ state of
${}^{171}\mathrm{Yb}^+$, shown in Fig.~\ref{fig:FStateZoom}.  The
qubit states can be defined as $\ket{0} \equiv |F,M_F\rangle =
\ket{3,0}$ and $\ket{1}\equiv \ket{3,1}$, and the auxiliary states as
$|A^{\mbox{\tiny(+)}} \rangle\equiv \ket{4,0}$ and
$|A^{\mbox{\tiny(-)}} \rangle \equiv \ket{4,1}$.  A stable, static
magnetic field provides the qubit splitting, and the qubit and
auxiliary manifolds are separated by the ${}^2\!\!\,F_{7/2}^o$
hyperfine splitting ($\approx \! 3.6 \mbox{ GHz}$; we will refer to
the coupling fields as microwaves).  Further, we require that the
system possess a means by which projective quantum measurement can be
performed selectively for population in each of these two
manifolds. In this example, detection can be effected by
hyperfine-selective transfer to the ground ${}^2\!\!\,S_{1/2}$ state
via optical pumping at $\lambda\!=\!760\mbox{ nm}$ on
${}^1\![3/2]_{3/2}^o \! \leftarrow \! {}^2\!\!\,F_{7/2}^o$, followed
by spontaneous emission on ${}^1\![3/2]_{3/2}^o \! \leadsto \!
{}^2\!\!\,S_{1/2}$. We have confirmed experimentally that this
measurement can be accomplished in a few milliseconds with greater
than 95\% hyperfine manifold selectivity \cite{ConradPC}, and the
theoretical limit is greater than $1- 10^{-5} $. 

Without loss of generality, we adopt a state vector description of the
gate operation for clarity. Before we describe the certified gate
protocol, we can consider the action of a general, unitary,
single-qubit gate $U(\mathbf{\hat{n}},\Theta) \equiv \exp\left(-\ii
\Theta \,\mathbf{\hat{n}} \!\cdot\! \boldgreek{\sigma}/2 \right) $ on
an arbitrary pure input state $\ket{\psi_0} \equiv \alpha \ket{0} +
\beta \ket{1}$.  If we rewrite the initial state in the basis of
$\ket{\pm\mathrm{n}}$ (the $\pm$ eigenvectors of $\mathbf{\hat{n}}
\cdot \boldgreek{\sigma}$), we have 
\begin{equation}
   \ket{\psi_0} = c^{\tiny{(+)}} \ket{+\mathrm{n}} + c^{\tiny{(-)}}
   \ket{-\mathrm{n}},
\end{equation}
where $c^{\tiny{(\pm)}} \equiv \langle \pm \mathrm{n} \ket{\psi_0}$.
The states $\ket{\pm\mathrm{n}}$ can likewise be written in terms of
the polar ($\theta$) and azimuthal ($\phi$) angles of
$\mathbf{\hat{n}}$ on the Bloch sphere as
\begin{eqnarray}
    \ket{ + \mathrm{n}} &=& \cos \left( \frac{\theta}{2} \right)
    \ket{0} + \rme^{\sii \phi} \sin \left( \frac{\theta}{2} \right)
    \ket{1} \nonumber \\
    \ket{ - \mathrm{n}} &=& \sin \left( \frac{\theta}{2} \right)
    \ket{0} - \rme^{\sii \phi} \cos \left( \frac{\theta}{2} \right)
    \ket{1}.\label{nbasis}
\end{eqnarray}
This choice of basis simplifies the expression describing the effect
of the gate to
\begin{equation}
    U(\mathbf{\hat{n}},\Theta) \ket{\psi_0} = \rme^{-\sii
      \frac{\Theta}{2}}c^{\tiny{(+)}} \ket{+\mathrm{n}} + \rme^{\sii
      \frac{\Theta}{2}} c^{\tiny{(-)}} \ket{-\mathrm{n}}.
\end{equation}

For a certifiable version of the gate $U(\mathbf{\hat{n}},\Theta)$,
first, a microwave pulse with four simultaneous tones ($f_i$, see
Fig.~\ref{fig:FStateZoom}) transfers (ideally all) the population from
the qubit states to the auxiliary states according to $\ket{\pm
  \mathrm{n}} \! \rightarrow \! |A^{\mbox{\tiny($\pm$)}} \rangle $.
Each of the $\ket{\pm \mathrm{n}}$ basis states is paired with only
one of the auxilliary states $|A^{\mbox{\tiny($\pm$)}} \rangle$ by two
of the four tones $f_i$ and acts as a coherent dark state with respect
to the other two.  The relative phases ($\varphi_{12}$ and
$\varphi_{34}$) and Rabi frequencies ($\Omega_i$) chosen for the four
frequencies depend only on the angles used to describe
$\mathbf{\hat{n}}$, $\phi$ and $\theta$ (respectively, see
Eq.~(\ref{nbasis})).  Specifically, $\varphi_{12} = \varphi_{3,4} =
\phi$, $\Omega_1 = \Omega_4 = \Omega \cos(\theta/2)$, and
$\Omega_2=\Omega_3 = \Omega \sin(\theta/2)$. In the rotating frame
with respect to the four splittings, the interaction Hamiltonian is
\begin{eqnarray}
H &=& |A^{\tiny{(+)}} \rangle \left( \frac{\Omega_1}{2} \bra{0} +
\frac{\Omega_2}{2} \rme^{-\sii \phi} \bra{1} \right) \nonumber \\
&& + |A^{\tiny{(-)}} \rangle \left( \frac{\Omega_3}{2} \bra{0} +
\frac{\Omega_4}{2} \rme^{-\sii (\phi + \pi)} \bra{1} \right) +
\mathrm{H.c.} \\
& = & \frac{\Omega}{2} \left( |A^{\tiny{(+)}}\rangle
\langle+\mathrm{n}| + |A^{\tiny{(-)}}\rangle \langle-\mathrm{n}| +
\mathrm{H.c.} \right) 
\end{eqnarray}
where we assume the splittings are such that the four frequencies are
non-degenerate.

Since these four sinusoids can be generated by a single synthesizer
(for instance, a digital arbitrary waveform generator utilizing a
single voltage reference) and can be made to share a single
transmission system, amplifier chain, antenna, etc., we consider the
case in which the amplitude error of this step is a fractional
amplitude error that is shared by all four coupling terms.  Since we
seek full transfer from the qubit manifold to the auxiliary manifold,
we represent the pulse area as $\int \! \mathrm{d}t \Omega = \pi +
\delta \pi_n$, where $\delta \pi_n$ is the result of an amplitude
error for the $n$th step of the gate. We can write the state of the
system after the (possibly imperfect) transfer as 
\begin{equation}
    \ket{\phi_1} = -\ii\! \cos(\frac{\delta\pi_1}{2}) \left(
    c^{\mbox{\tiny(+)}} |A^{\mbox{\tiny(+)}}\rangle +
    c^{\mbox{\tiny(-)}} |A^{\mbox{\tiny(-)}}\rangle \right) -
    \sin(\frac{\delta \pi_1}{2}) \ket{\psi_0}
\end{equation}
which is in the desired form for error detection,
\begin{equation}
    \ket{\phi_{n+1}} =  \sqrt{1-|\epsilon|^2} \,\ket{\psi_{n+1}} +
    \epsilon \ket{\psi_n}, \label{eq:DesiredErrorForm}
\end{equation}
if we identify the error as $\epsilon = -\sin(\delta \pi_1/2)$.

Next, any population left in the qubit manifold ($\ket{0}$ and
$\ket{1}$, see Fig.~\ref{fig:FStateZoom}) is dissipatively transferred
to ${}^2\!\!\;S_{1/2}$ via optical pumping.  This ``clean out''
process will be accompanied by subsequent fluorescence detection of
ground-state population at some point -- right away or potentially
even up until very end of an algorithm.  If the ion is queried
immediately, it will yield fluorescence (a ``bright state ion'') with
small probability $\sin^2(\delta \pi_1/2)$.  If the ion is not in the
bright state, the dissipative process has completed the successful
transfer of all qubit population to the auxiliary manifold, yielding
the desired target state free of that error, $\ket{\psi_1} = -\ii
(c^{\mbox{\tiny(+)}} |A^{\mbox{\tiny(+)}} \rangle +
c^{\mbox{\tiny(-)}} |A^{\mbox{\tiny(-)}} \rangle$).

For the third step, a second pulse with the same four tones is applied
to transfer (ideally all) the population from the auxiliary manifold
back to the qubit manifold.  The only difference between the waveform
for the first and second pulses is that a common phase shift
$\pi-\Theta/2$ is added to tones $f_1$ and $f_2$ only, and a common
phase shift of $\pi+\Theta/2$ is added to tones $f_3$ and $f_4$ only.
Again keeping track of a potential (possibly different) amplitude
error that gives rise to finite $\delta \pi_2$ in the nominal
$\pi$-pulse, the system is left in
\begin{eqnarray}
    \ket{\phi_2} &=& \cos(\frac{\delta \pi_2}{2}) \left(\rme^{-\sii
      \frac{\Theta}{2}}c^{\mbox{\tiny(+)}} \ket{+\mathrm{n}} +
    \rme^{\sii \frac{\Theta}{2}} c^{\mbox{\tiny(-)}} \ket{-\mathrm{n}}
    \right) \nonumber \\
    && - \! \sin(\frac{\delta \pi_2}{2}) \ket{\psi_1},
\end{eqnarray}
which is in form (\ref{eq:DesiredErrorForm}) for $\epsilon = -
\sin(\delta\pi_2/2)$.

As the final step, any population left in the auxiliary manifold
($|A^{\tiny(\pm)}\rangle$, viz.~$\ket{\psi_1}$) is optically pumped to
the ground state, either yielding a bright state (with probability
$\sin^2(\delta\pi_2/2)$) or completing the transfer to produce
\begin{equation}
    \ket{\psi_2} = U(\mathbf{\hat{n}},\Theta) \ket{\psi_0} =
    \rme^{-\sii \frac{\Theta}{2}}c^{\mbox{\tiny(+)}} \ket{+\mathrm{n}}
    +  \rme^{\sii \frac{ \Theta}{2}}c^{\mbox{\tiny(-)}}
    \ket{-\mathrm{n}}, \label{eq:PsiFinalCoupled}
\end{equation}
the ideal gate with no contribution from the amplitude errors.

The gate protocol above provides a means for certifying the operation
against fractional amplitude errors that are shared by the four tones
in either of the two pulses.  With respect to this error model,
whether we check for a bright state immediately or delay the flag
query, the dissipative transfer of leftover population to the bright
state either leaves the ion in the bright state or accomplishes
errorless operation of the gate.  The overall probability of
error-free operation is
$(1-\sin^2(\delta\pi_1/2))(1-\sin^2(\delta\pi_2/2)) \approx 1- 2
(\overline{\delta \pi}/2)^2$ (where $\overline{\delta \pi}$ is an
average error during this sequence) and for uncorrelated errors, the
overall error probability is $\sqrt{2}$ larger than the case without
the out-coupling for error detection ($\approx \sqrt{2}
(\overline{\delta \pi}/2)^2$).  For single, isolated gates, this
accomplishes no error correction, but the error detection can be used
as a means to select instances that are trustworthy against this type
of error.  For instance, the high-quality rotations that are required
to perform quantum state or process tomography could be certified
against conflating errors in the state/process with this type of error
introduced by the tomography process.  Perhaps more importantly, more
trustworthy NISQ-era \cite{Preskill2018quantum} computational results
can be sorted from those that are flagged by this process as
containing errors, which may prove a useful way to assess the
confidence of a result.

The gate certification idea above is also extendable to multi-qubit
gates and other types of errors.  Next, we consider two examples of
particularly troublesome error sources in the trapped ion platform:
qubit addressing errors, and errors in 2-qubit entangling gates. 

For trapped ions with hyperfine qubits, an addressed single-qubit gate
can be driven by a focused laser beam where the ``microwave'' signals
are actually in optical beatnotes that drive stimulated Raman
transitions.  If the first step of the certified single-qubit gate
described earlier is applied to ion $j$ by one such laser beam, there
can be a non-negligible amount of light that illuminates neighboring
ions and moves a small amount of their qubit populations to their
auxiliary manifolds.  To deal with this, the optical pumping beam
addressed to ion $j$  can be augmented by a series of optical pumping
beams on the neighboring ions (or further) that are tuned to clean out
those ions' auxiliary manifolds.  This will either flag an addressing
error by producing a bright state, or, more likely, undo any errant
transfer from imperfect addressing by the stimulated Raman beam.  The
same process can then be applied for the second half of the gate being
run on ion $j$, except that now the clean out will have all optical
pumping beams (including $j$) set to clean out the auxiliary
manifolds.   Addressing errors of the optical pumping beams themselves
are still possible in the first half of the gate, but these will also
be flagged by the appearance of a bright state. A lack of bright state
qubits, therefore, certifies the gate against both Raman beam and
optical pumping beam addressing errors -- if all ions are found to be
dark, these addressing errors have been eliminated.

For multi-qubit gates, we choose as an example the Cirac-Zoller (CZ)
gate \cite{Cirac1995Quantum} since it maps easily onto a series of
discrete population transfer steps.  We consider two ions ($m$ and
$n$) in arbitrary initial qubit states and one motional mode of
frequency $\nu$ prepared in its ground state.  Frequency selectivity
can be used to drive ``carrier'' ($\sigma_+ + \sigma_-$), ``red
sideband'' ($a\sigma_+ + a^\dagger \sigma_-$), or ``blue sideband''
($a^\dagger \sigma_+ + a \sigma_-$) transitions, where $\sigma_\pm$
and $a^\dagger,a$ are the atomic and motional raising and lowering
operators, respectively.  To be consistent with the original proposal
by Cirac and Zoller and to avoid confusion with motional Fock state
labels, we adopt $\ket{\mathrm{e}_{m/n}}$ and $\ket{\mathrm{g}_{m/n}}$
as the notation for the qubit states.

\begin{figure}
\includegraphics[width=0.45\textwidth]{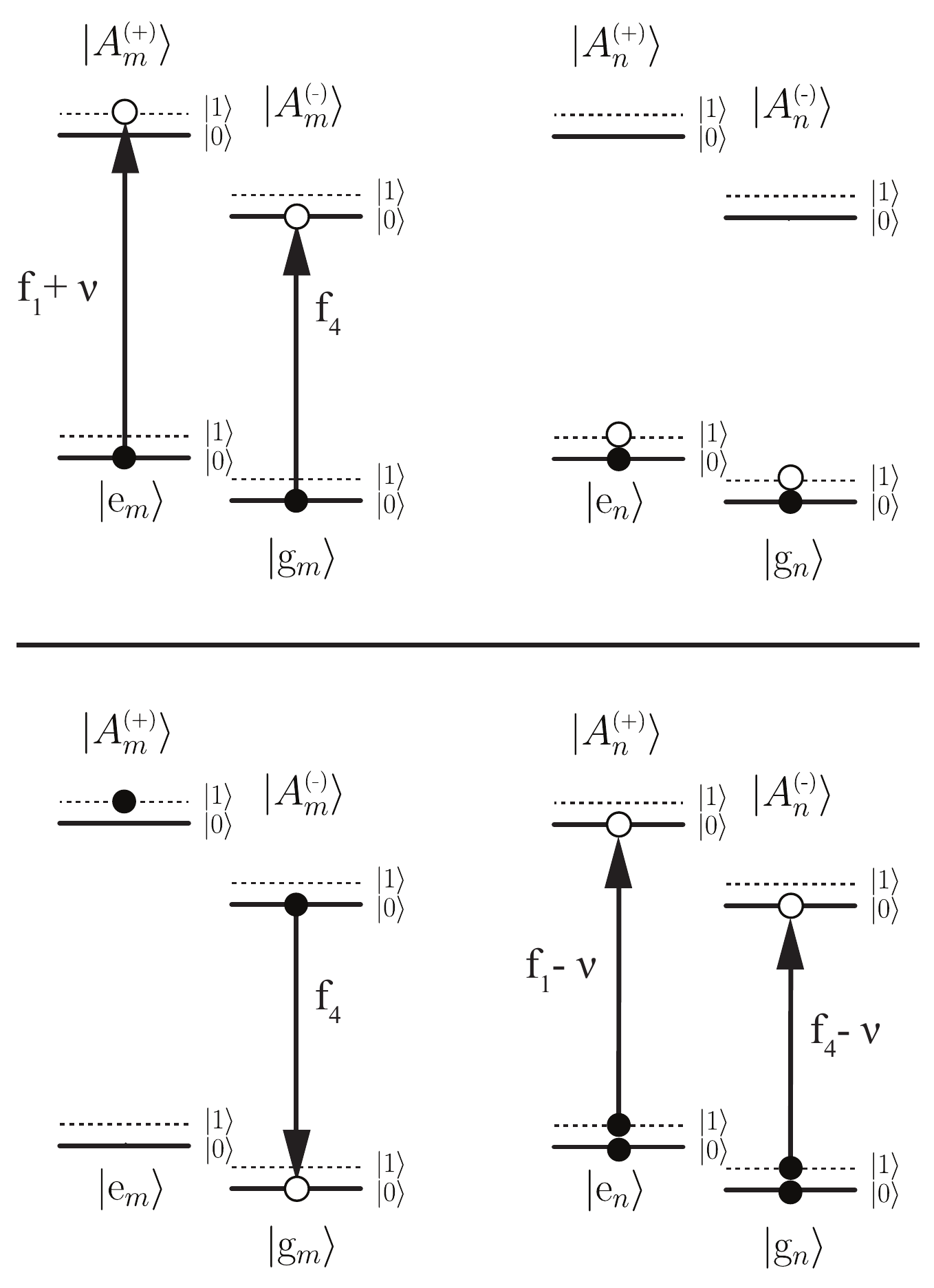}
\caption{Protocol for a certified Cirac-Zoller entangling gate between
  ion $m$ (left) and ion $n$ (right). The top panel shows the first
  transfer, which would be followed by optical pumping of any
  remaining population in the qubit manifold of ion $m$.  The second
  transfer is shown in the bottom panel, which would also be followed
  by state-selective opitcal pumping of leftover population to the
  ${}^2\!\!\,S_{1/2}$ state.  Transfer 3 is the reverse of the lower
  panel (with a phase shift of $\pi$ added to $f_1 - \nu$), and
  transfer 4 is the reverse of the upper panel.}
\label{fig:CZLevels}
\end{figure}

A certifiable version of the CZ gate proceeds in four transfers, shown
in Fig.~\ref{fig:CZLevels}. We start with an initial (possibly
entangled) state $\ket{\psi_0} =
(c_\mathrm{ee}\ket{\mathrm{e}_m,\mathrm{e}_n}+c_\mathrm{ge}\ket{\mathrm{g}_m,\mathrm{e}_n}
+ c_\mathrm{eg}\ket{\mathrm{e}_m,\mathrm{g}_n}+
c_\mathrm{gg}\ket{\mathrm{g}_m,\mathrm{g}_n}) \otimes \ket{0}$.
First, two simultaneous tones (on a non-copropagating stimulated Raman
beam) are applied to only ion $m$ such that $f_1 + \nu$ transfers
$\ket{\mathrm{e}_m,0} \! \rightarrow \! \ket{A^{\mbox{\tiny(+)}}_m,1}$
(i.e.~on a blue sideband) and $f_4$ drives $\ket{\mathrm{g}_m,0} \!
\rightarrow \! \ket{A^{\mbox{\tiny(-)}}_m,0}$ (carrier).  This is
followed by hyperfine-resolved optical pumping of any residual
population in qubit manifold of ion $m$ to ${}^2\!\!\,S_{1/2}$. If ion
$m$ is not found in the bright state after this process (the
probability of which is $1-\sin^2(\delta \pi_1/2)$), the system will
be left in the state 
\begin{eqnarray}
 \ket{\psi_1} & = & -\ii ( c_\mathrm{ee}
 \ket{A^{\mbox{\tiny(+)}}_m,\mathrm{e}_n,1} + c_\mathrm{eg}
 \ket{A^{\mbox{\tiny(+)}}_m,\mathrm{g}_n,1}  \nonumber \\
&&  + c_\mathrm{ge} \ket{A^{\mbox{\tiny(-)}}_m,\mathrm{e}_n,0} +
 c_\mathrm{gg} \ket{A^{\mbox{\tiny(-)}}_m,\mathrm{g}_n,0} ).
\end{eqnarray}

Second, three frequencies will be applied simultaneously and
individually addressed as follows.  Ion $m$ will be driven with $f_4$
to transfer $\ket{A^{\mbox{\tiny(-)}}_m,0} \! \rightarrow \!
\ket{\mathrm{g}_m,0}$, while $f_1 - \nu$ and $f_4 - \nu$ are applied
to ion $n$ to drive sidebands $\ket{\mathrm{e}_n,1} \! \rightarrow \!
\ket{A^{\mbox{\tiny(+)}}_n,0}$ and $\ket{\mathrm{g}_n,1} \!
\rightarrow \! \ket{A^{\mbox{\tiny(-)}}_n,0}$.  After this transfer,
assuming it is imperfectly executed due to a pulse area error shared
by all three tones, the system is in the state
\begin{eqnarray}
    \ket{\phi_2} = \cos(\frac{\delta\pi_2}{2}) \ket{\psi_2} -
    \sin(\frac{\delta\pi_2}{2}) \ket{\psi_1}, \label{eq:CZphi2}
\end{eqnarray}
where 
\begin{eqnarray}
 \ket{\psi_2} &  \equiv & - \left(c_\mathrm{gg}
 \ket{\mathrm{g}_m,\mathrm{g}_n} + c_\mathrm{ge}
 \ket{\mathrm{g}_m,\mathrm{e}_n} \right. \nonumber \\
&& \left. + c_\mathrm{eg}
 \ket{A^{\mbox{\tiny(+)}}_m,A^{\mbox{\tiny(-)}}_n} + c_\mathrm{ee}
 \ket{A^{\mbox{\tiny(+)}}_m,A^{\mbox{\tiny(+)}}_n} \right) \otimes
 \ket{0},
\end{eqnarray}
and which is in form (\ref{eq:DesiredErrorForm}) with $\epsilon =
-\sin(\delta \pi_2/2)$.

At this point, the residual populations that need to be optically
pumped to the ground state ($\ket{A^{\mbox{\tiny(-)}}_m,0}$,
$\ket{\mathrm{g}_n,1}$, and $\ket{\mathrm{e}_n,1}$) are close in
energy to populated levels.  If the resolution of the optical pumping
step is sufficient for this, they can be cleaned out directly; if not,
a multi-step transfer-then-pump process involving additional
resolvable auxiliary states may be required.  In either case, once
this step has been completed (and assuming neither ion is in the
bright state), the system is left in $\ket{\psi_2}$.

The next step is almost identical to the preceding one: $f_4$ is
applied to ion $m$ and $f_1 - \nu$ and $f_4 - \nu$ are applied to ion
$n$.  However, a phase shift of $\pi$ is added to tone $f_1 - \nu$
(which is on ion $n$ only) for this transfer.  Once the transfer
attempt is completed, clean out of the auxiliary manifold of ion $n$
and the $\ket{\mathrm{g}_m}$ level of ion $m$ will leave the system in 
\begin{eqnarray}
 \ket{\psi_3} &  = &  \ii c_\mathrm{gg}
 \ket{A^{\mbox{\tiny(-)}}_m,\mathrm{g}_n,0} +  \ii c_\mathrm{ge}
 \ket{A^{\mbox{\tiny(-)}}_m,\mathrm{e}_n,0} \nonumber \\
&& +\ii c_\mathrm{eg} \ket{A^{\mbox{\tiny(+)}}_m,\mathrm{g}_n,1} - \ii
 c_\mathrm{ee} \ket{A^{\mbox{\tiny(+)}}_m,\mathrm{e}_n,1}.
\end{eqnarray}
The last step is the same as the first, but with the subsequent
clean-out occurring on the auxiliary manifold of ion $m$.  Again
conditioned on the fact that the ions are not in the bright state, the
motion factors and the final state of the two-ion system is
\begin{eqnarray}
   \ket{\psi_4} & =& U_\mathrm{CZ} \ket{\psi_0} \nonumber \\
    &= & ( c_\mathrm{gg} \ket{\mathrm{g}_m,\mathrm{g}_n} +
   c_\mathrm{ge} \ket{\mathrm{g}_m,\mathrm{e}_n} \nonumber \\
&& + c_\mathrm{eg} \ket{\mathrm{e}_m,\mathrm{g}_n} - c_\mathrm{ee}
   \ket{\mathrm{e}_m,\mathrm{e}_n} ) \otimes \ket{0}
   .\label{eq:FinalCZState}
\end{eqnarray}

The state in Eq.~(\ref{eq:FinalCZState}) is identical to the result of
an ideally executed the CZ gate, which can perform a CNOT gate with
the addition of some single-qubit rotations \cite{Cirac1995Quantum}.
However, each step can now be checked for errors that affect the pulse
area of the simultaneous tones the same way, such as a drift in signal
strength.  Upon passing the check, subsequent gates will not be
susceptible to correlated error accumulation from the errors removed
by this scheme.

We have introduced here methods that allow certification of all of the
gates required for universal quantum computation against common-mode
pulse area errors. However, the classes of errors that are encountered
in implementations of quantum processors go well beyond the limited
class addressed here, as do the details for how these systems are made
to execute their universal set of gates.  The general approach we have
sketched for developing certification should be adaptable to some of
these situations, and extensions of these ideas may be possible moving
forward.

\textit{Acknowledgments }WCC acknowledges helpful discussions with
Dave Hayes, Eric Hudson, Paul Hamilton, and David Hucul. This work was
supported by the U.S.~Army Research Office under Grant
No.~W911NF-19-S-0011 and the U.S.~National Science Foundation under
Award No.~PHY-1912555.

\bibliography{StabilizedGates}
\end{document}